# Single-Spin Observables and Orbital Structures in Hadronic Distributions

Dennis Sivers

Portland Physics Institute
4730 SW Macadam, Suite 101
Portland, Oregon 97239

Spin Physics Center
University of Michigan
Ann Arbor, Michigan 48109-1120

## ABSTRACT

Single-spin observables in scattering processes (either analyzing powers or polarizations) are highly constrained by rotational invariance and finite symmetries. For example, it is possible to demonstrate that all single-spin observables are <u>odd</u> under the finite transformation $O = PA_\tau$ where P is parity and $A_\tau$ is a finite symmetry that can be designated "artificial time reversal". The operators P, O and $A_\tau$ all have eigenvalues $\pm 1$ so that all single-spin observables can be classified into two distinct categories:

1. P-odd and $A_\tau$ - even
2. P-even and $A_\tau$ - odd.

Within the light-quark sector of the standard model, P-odd observables are generated from point-like electroweak processes while $A_\tau$ - odd observables (neglecting quark mass parameters) come from dynamic spin-orbit correlations within hadrons or within larger composite systems, such as nuclei. The effects of $A_\tau$-odd dynamics can be inserted into transverse-momentum dependent constituent distribution functions and, in this paper, we construct the contribution from an orbital quark to the $A_\tau$ –odd quark parton distribution $\Delta^N G_{q/p\uparrow}^{front}(x, k_{TN}; \mu^2)$. Using this distribution, we examine the crucial role of initial- and final-state interactions in the observation of the scattering asymmetries in different hard-scattering processes. This construction provides a geometrical and dynamical interpretation of the Collins conjugation relation between single-spin asymmetries in semi-inclusive deep inelastic scattering and the asymmetries in Drell-Yan production. Finally, our construction allows us to display a significant difference between the calculation of a spin asymmetry generated by a hard scattering mechanism involving color-singlet exchange (such as a photon) and a calculation of an asymmetry with a hard-scattering exchange involving gluons. This leads to an appreciation of the process dependence inherent in measurements of single-spin observables.

## I. Introduction

The theoretical calculation of single-spin observables presents some interesting technical issues for the standard model. For many years, it was argued in lectures and in conferences [1] that QCD "predicts" that single-spin asymmetries in hard-scattering processes such as $pp\uparrow \to jetX$ or $ep\uparrow \to ehX$ must vanish. This argument can be traced to an early approach [2] to the factorization of the QCD hard-scattering parton model that associated the spin observable with an underlying single-spin asymmetry at the quark level,

$$A_N(qq\uparrow \to qX) = \alpha_s \frac{m_q}{\sqrt{s}} f(\theta), \tag{1.1}$$

and then convoluted the result with the quark transversity distribution using the assumption of collinear factorization. For light quarks this approach does lead to a vanishingly small asymmetry. However, the perturbative asymmetry (1.1) contains the ratio $m_q/\sqrt{s}$, and the result of this approach can therefore be classified as a "higher-twist" observable. Many other types of "higher-twist" mechanisms exist within the QCD hard-scattering model and it is possible that some of these mechanisms can overpower the asymmetry generated from eq. (1.1). Therefore, there is no prediction of small asymmetries within the general framework of collinear factorization of QCD but merely a challenge to find possible mechanisms and to use them to make predictive calculations.

There are two basic approaches to classifying the potential mechanisms for single-spin observables. The first is based on the use of $k_T$-dependent parton distribution functions [3] or parton fragmentation functions [4] within the hard-scattering model. The second is based directly on the twist expansion of the operator product formalism [5] Roughly speaking, the distinction between these two approaches can be compared to differences between the Schrodinger (wave-function) approach and the Heisenberg (operator) approach to classical quantum mechanics. Each can provide a systematic framework for the separation of the perturbatively-calculable short-distance component of a physical observable from the "soft" nonperturbative dynamics associated with composite systems. This paper will stick with the concept of $k_T$-dependent distribution functions because it provides the opportunity to directly parameterize orbital angular momentum. For a proton polarized in the y-direction (normal to the scattering plane) we introduce a quark with orbital angular momentum in the y-direction and then calculate the contribution to inclusive scattering associated with scattering from the orbiting quark.

For convenience, the conventions of Jacob and Wick [6] in scattering processes are described in Fig. 1 while Table 1 gives the behavior of the different components of a single-spin observable under the finite transformations P (parity), C (charge conjugation), T (time reversal) and $A_\tau$ (artificial time reversal). We will be describing an $A_\tau$- odd observable that results from the preferential scattering from one or more segments or the rotating quark's orbit within the polarized proton. In order to allow for the possibility that one part of the orbit can contribute more than another we show that the presence of initial- state and/or final-state interactions can screen or modify the kinematics of the underlying hard-scattering process. [7]

The essential contribution of these initial- and final-state interactions to the existence of an observable asymmetry arising from an orbital distribution defines a fundamental challenge to formalism of describing hard-scattering processes. This challenge was not adequately presented in ref. 3 where it was merely asserted that spin-orbit correlations allow for the existence of the corresponding distribution. The challenge is now being addressed by spectator models [8] for orbital distributions and by other approaches [9] that illustrate specific aspects of the underlying dynamics. However, it is also possible to demonstrate the role of spin-orbit effects in scattering processes by explicit construction. The construction presented here focuses on the symmetries of the composite system and on the universal structure of an orbiting system described by its orbit parameters. The process-dependent soft interactions are crucial components of the hard-scattering formalism that determine how the orbital structure is probed. In some processes, the effect of the crucial initial- or final-state interactions can be absorbed into an "eikonal-charge" measuring the screening effects. In other processes, spin-oriented binding effects lead to spin-oriented initial- or final-state effects that contribute directly to the asymmetry. The factorization properties of the hard-scattering model must be understood in a manner that accounts for these different contributions.

The outline of this paper is as follows. Because there has been consistent confusion about the constraints of finite symmetries in quantum theories as applied to single-spin observables, Sec. II presents an introductory discussion to explain the distinction between time reversal (T) and "artificial time reversal" ($A_\tau$). This section also discusses the use of $A_\tau$ to form projection operators that can isolate mechanisms leading to parity-conserving single-spin asymmetries. These projection operators function either at the amplitude level or at the level of cross sections. They have proven to be a useful tool to aid in the construction of a partonic description of $A_\tau$-odd observables or to enable specific calculations. Sec. III discusses a local gauge-invariant quantum mechanical description of the scattering from a rotating constituent within a composite system such as a hadron. It presents an explicit construction of the $A_\tau$-odd distribution $\Delta^N G_{q/p\uparrow}^{front}(x, k_{TN}; \mu^2)$ that fully describes an orbiting quark within a polarized proton. The construction gives the normalization of the distribution in terms of the quark's mean orbital angular momentum and displays some of the function's basic symmetries. Sec. IV then gives a brief presentation of the role of initial- and final-state interactions in the scattering process. It is pointed out that these interactions are necessarily present both in the spin-dependent and in the spin-averaged processes. In spin-averaged scattering processes they are absorbed into the "measured" constituent distributions. For single-spin asymmetries, they can also be absorbed into the distributions but they play a more direct role in that they completely determine the sign and the magnitude of the observed effects. We explain how the Collins [10] symmetry property,

$$\Delta^N G_{q/p\uparrow}(Drell-Yan) = -\Delta^N G_{q/p\uparrow}(SIDIS), \qquad (1.2)$$

takes on a geometrical and dynamical meaning by relating the initial-state interactions in the Drell-Yan process to the final-state interactions in semi-inclusive deep inelastic scattering using charge conjugation. This approach is complementary to the original derivation by Collins in Ref. 10 and serves to further illustrate the fundamental nature of this conjugation relation. We also show that there is a significant difference between a process with a hard-scattering mechanism that involves a color-singlet exchange (such as a photon) and one that involves gluon exchange. In photon processes the struck quark does not change color and remains subject to a spin-oriented confining force. This force then produces the spin-oriented lensing effect that has been identified in the model of Burkardt [9] for single-spin asymmetries in semi-inclusive deep inelastic scattering. By contrast, in gluon-exchange processes the struck quark is liberated from the spin-oriented confining force by the hard-scattering mechanism. The comparison demonstrates both the high potential interest in different experimental methods and the importance of understanding the process dependence in the measurement of orbital structure functions. Sec. V concludes with some simple observations about coherent effects in composite systems and discusses the prospects for using the process-dependent twist-3 operators to augment calculations with transverse momentum dependent structure functions. The contents of Sec. V serve to define some unsolved problems. This paper does not address questions involving the dynamical origin of spin-orbit effects in hadrons but defers these important issues for a future article.

**II. Time Reversal and Artificial Time Reversal.**

Many articles on single-spin asymmetries written in the mid 1990's refer to "T-odd" parton distribution functions or parton fragmentation functions. It is difficult to trace exactly how this unfortunate language became acceptable since the designation was always applied to tensor products that were explicitly even under the time reflection operator. It is sometimes clear from context that an author who adopted the language was aware that the term "T-odd" did not refer to time reversal but to a transformation using the linear operator $A_\tau$ described in the introduction. The terms "naïve" time reversal and "T but not CP" were occasionally used to describe the operator that I have here labelled

"artificial time reversal." Frequently, however, the confusion of language led to incorrect statements or conclusions. In order to alert the reader to these historical problems and to clarify the useful properties of the $A_\tau$ operator we will review here some of the elementary properties of the time reversal operator in classical mechanics, quantum mechanics and quantum field theory. This will not be a thorough discussion such as can be found in many textbooks [11] but will be aimed at illustrating the role of the $A_\tau$ operator in projecting transversity amplitudes [12], and in organizing calculations of single-spin observables. By clarifying the distinction between time reversal and artificial time reversal we recover the ability to invoke CPT invariance and crossing relations to relate different processes and are able to use charge conjugation ( C ) and G-parity (charge conjugation followed by a rotation in isospin space) to help understand the dynamics associated with the quantum structures that generate single-spin observables.

In the remainder of this section we will frequently use lower case Latin subscripts to denote 3-dimensional vectors. In this notation, the time reversal invariance of Newtonian mechanics is easily displayed, and we can see that Newton's equations,

$$F_i = ma_i, \qquad (2.1)$$

involve only second-order time derivatives,

$$a_i = \frac{d^2}{dt^2} x_i, \qquad (2.2)$$

so that the transformation $t \to -t$ leaves the equations unchanged. There are no mechanisms that violate time reversal invariance in Newtonian mechanics.

There are, however, some interesting challenges in constructing a time reflection operator, T, in quantum mechanics. Consider the nonrelativistic Schrodinger equation in the form

$$i \frac{\partial}{\partial t} \Psi(x,t) = H \Psi(x,t) \qquad (2.3)$$

where H is the Hamiltonian operator. We understand this equation to mean that $\Psi(x,t)$ runs forward in time. That is, if we specify $\Psi(x,t_0)$, then eq. (2.3) gives the behavior of $\Psi(x,t)$ for $t \geq t_0$. If we take the complex conjugation of eq. (2.3)

$$-i \frac{\partial}{\partial t} \Psi^*(x,t) = H^* \Psi^*(x,t)$$

$$i \frac{\partial}{\partial (-t)} \Psi^*(x,t) = H^* \Psi^*(x,t) \qquad (2.4)$$

we have created an equation that runs backward in time so that specifying $\Psi^*(x,t_0)$ and solving the equation gives $\Psi^*(x,t)$ for $t \leq t_0$. This simple example illustrates the important observation that all statements about the time reversal operator in quantum theory are necessarily <u>representation dependent</u> in that both the boundary conditions of the quantum system and the operation of complex conjugation have to be explained explicitly. It also demonstrates that T cannot be a unitary operator as it is not linear. Instead, T can be decomposed into two parts,

$$T = U\mathrm{K} \tag{2.5}$$

where U is unitary and K is an antilinear operator that plays the role of the complex conjugation operator in (2.4). For this reason, T is sometimes called <u>anti-unitary</u>. The anti-unitary nature of T provides a clue that any quantum representation of time reversal requires some vigilance to achieve consistent interpretation of the action of the operator.

Consider now, the Heisenberg picture in which the time variation of an operator Q is given by the equation of motion

$$i\frac{\partial Q}{\partial t} = [Q, H] \tag{2.6}$$

where, again, H denotes the Hamiltonian operator and the brackets represent the commutator. We consider the effect of time reversal on this equation

$$-i\frac{\partial (TQT^{-1})}{\partial t} = [TQT^{-1}, THT^{-1}] \tag{2.7}$$

and note that invariance under time reversal requires that the Hamiltonian operator must commute with T,

$$THT^{-1} = H; TH = HT. \tag{2.8}$$

This implies that any numerical coefficients in H are real and that any operators contained in H can be combined to form scalars under T. Rewriting (2.7) in the form ($\overline{Q} = TQT^{-1}$)

$$i\frac{\partial \overline{Q}}{\partial (-t)} = [\overline{Q}, H] \tag{2.9}$$

shows that $\overline{Q}$ has the same equation of motion running backwards in time as does Q running forward in time. In this operator formulation of quantum systems the expectation values of Q and $\overline{Q}$ are often simply related. For example if Q is the position operator, $Q = x_i$, of a particle then we can have (depending on initial conditions)

$$\langle Q \rangle = \langle \overline{Q} \rangle \tag{2.10}$$

with $\overline{Q}$ traversing the path of Q in reverse. If Q denotes the momentum operator, $p_i$, the angular momentum operator, $J_i$, the spin operator, $\sigma_i$, or the orbital angular momentum operator, $L_i$, depending on boundary conditions, we can have

$$\langle Q \rangle = -\langle \overline{Q} \rangle \tag{2.11}$$

so that the overall behavior of the classical limit is preserved. Operators obeying (2.11) can be said to be odd under T. With the requirement that it be invariant under time reversal, we see that the Hamiltonian cannot contain terms linear in operators that are odd under time reversal, such as $p_i$ or $L_i$. Also, it cannot contain any scalar products, such as $x_i J_i$ that would change sign under time reversal. However, scalar products such as $p_i p_i$ or (importantly for our discussion) $L_i \sigma_i$ can be accommodated within the Hamiltonian. Specifically, spin-orbit terms are allowed by time reflection invariance and it is precisely these dynamical effects that we are going to use to describe single-spin asymmetries in hard scattering processes in the QCD parton model.

In classical mechanics we are familiar with the fact that certain operations such as rotations do not commute. In quantum mechanics the number of noncommuting operators is larger and it is recognized that we have to commutation relations into account. Therefore, in the study of quantum systems we often deal with a product of operators where it is important to keep track of the order in which the products are formed because some of the quantum operators do not commute. In studying such ordered products of operators, we have to keep in mind that, under the time-reversal operator T, the order of the operators in the product must be reversed without consideration of their commutation relations. That is, for example

$$T(Q_1 Q_2 Q_3 Q_4 Q_5)T^{-1} = \overline{Q_5}\,\overline{Q_4}\,\overline{Q_3}\,\overline{Q_2}\,\overline{Q_1} \qquad (2.12)$$

This important property of the time reflection operator plays a fundamental role both in the formulation of the CPT theorem and in the framing of the Spin Statistics theorem in quantum field theory. [13] It is obviously a fundamental feature of what is meant by time reversal invariance. For our more modest purposes, we see that it explicitly plays an important role in scattering processes. Consider, again, the scattering process illustrated in Fig. 1. Assume, for simplicity, that all of the particles shown are spinless, but that particle 2 is part of extended system with angular momentum $J_i = J\hat{y}_i$ normal to the scattering plane. By convention [6] the normal to the scattering plane is defined by

$$\hat{y}_i = (\hat{k}_1 \times \hat{k}_3)_i / \sin \phi_{13} \qquad (2.13)$$

.

In the time-reversed process, the normal to the scattering plane is therefore defined by

(2.14)
$$T(\hat{y}_i)T^{-1} = ((-\hat{k}_3) \times (-\hat{k}_1))/\sin \phi_{31}$$
$$\hat{y}_{iT} = -\hat{y}_i$$

In conformance with the reversal of ordered products indicated in (2.12) the triple product

$$J_i(\hat{k}_1 \times \hat{k}_3)_i = \varepsilon_{ijk} J_i \hat{k}_{1j} \hat{k}_{3k} \qquad (2.15)$$

is therefore explicitly even under time reflection. Letting $J_i = \sigma_i$ we have

$$T(\hat{\sigma} \cdot (\hat{k}_1 \times \hat{k}_3))T^{-1} = \hat{\sigma} \cdot (\hat{k}_1 \times \hat{k}_3) \qquad (2.16)$$

This provides a simple example of how the anti-unitary nature of T requires attention to the complete set of conventions and not just the overall phase convention. Products of the form (2.16) are often erroneously called "T-odd" in papers dealing with single-spin asymmetries. Readers must be careful not to confuse them with operators such as $p_i, J_i, \sigma_i$ that are truly T-odd as defined in (2.11). Such confusion can lead to serious mistakes involving the application of time reversal invariance.

In contrast to the complications of the time-reversal operator, the operator $A_\tau$ mentioned in the introduction is a linear operator defined by the kinematic substitutions

$$A_\tau(k_i; \sigma_j) A_\tau^{-1} = (-k_i; -\sigma_j) \qquad (2.17)$$

for all particles. Notice the comparison with the parity operator, P, which generates the substitutions

$$P(k_i; \sigma_j) P^{-1} = (-k_i; \sigma_j) \qquad (2.18)$$

for all particles. Therefore, the product $O = PA_\tau$ has the property,

$$O(k_i; \sigma_j) O^{-1} = (k_i; -\sigma_j) \qquad (2.19)$$

for all particles. Equation (2.19) shows that all single-spin observables are necessarily odd under the combination $O = PA_\tau$. Equation (2.17) shows that the action of T and $A_\tau$ can be the same for an isolated, non-interacting, single particle system. This is the reason that $A_\tau$ can be designated "artificial time reversal". For any other dynamical system, the anti-unitarity property of T and the fact that $A_\tau$ does not alter the order of quantum operators as in Eq. (2.12) implies that their properties are quite different. In scattering theory, the application of the T operator requires the interchange of "in" and "out" states so that the ordering properties of the scattering matrix are preserved. The operator $A_\tau$ does not require this further action. The comparison between T and $A_\tau$ can be highlighted by

$$A_\tau(\hat{\sigma} \cdot (\hat{k}_1 \times \hat{k}_2)) A_\tau^{-1} = -\hat{\sigma} \cdot (\hat{k}_1 \times \hat{k}_2) \tag{2.20}$$

so by (2.16) and (2.20) this type of expression is <u>even</u> under T and <u>odd</u> under $A_\tau$ and O. The implications of rotational invariance plus finite symmetries for single spin observables are summarized in Table 1. Notice that the operators P, O and $A_\tau$ have a simple group structure defined by

$$PO = A_\tau$$
$$OA_\tau = P \tag{2.21}$$
$$A_\tau P = O$$

with

$$P^2 = O^2 = A_\tau^2 = 1$$
$$POA_\tau = 1 \tag{2.22}$$

These three operators are all very useful and we notice that O can be identified as a dual form of the parity operator that interchanges the roles of vectors and pseudovectors and that $A_\tau$ can be treated as a compound transformation. This group structure leads to the classification for single-spin observables into two distinct categories:

1. P-odd and $A_\tau$-even
2. P-even and $A_\tau$-odd.

We now turn to some of the basic properties of the $A_\tau$ operator that are useful in quantum calculations. As can be seen from the definition in Eq. (2.17)

$$A_\tau(\hat{k} \cdot \hat{\sigma}) A_\tau^{-1} = \hat{k} \cdot \hat{\sigma} \tag{2.23}$$

while

$$P(\hat{k} \cdot \hat{\sigma}) P^{-1} = -\hat{k} \cdot \hat{\sigma}$$
$$O(\hat{k} \cdot \hat{\sigma}) O^{-1} = -\hat{k} \cdot \hat{\sigma} \tag{2.24}$$

In QCD perturbation theory, P is conserved and the only $A_\tau$-odd effects are associated with quark mass parameters. Therefore, in the light-quark sector of the standard model where the u,d quark masses are neglected compared to $\Lambda_{qcd}$ there are no $A_\tau$-odd or O-odd effects that occur in QCD perturbation theory. The conservation of the $A_\tau$ operator in QCD perturbation theory with vanishing quark masses thus describes the content of the result of Kane, Pumplin and Repko [2] given in Eq. (1.1). There are, however, no requirements for $A_\tau$ to be conserved by long-range coherent forces such as those associated with color confinement or chiral symmetry breaking and, using parity conservation, we can show that QCD effects odd under $A_\tau$ are uniquely associated with coherent spin-orbit correlations.  These operators allow for the definition of a spin-oriented momentum that makes a local description of spin-orbit effects possible. Therefore, in the standard model, P-odd single spin observables are generated from point-like electroweak processes while $A_\tau$-odd spin observables are associated either with quark mass parameters or coherent spin-orbit effects. Thus, one of the most useful properties of the $A_\tau$ operator involves the application of the projection operators

$$P_A^{\pm} = (\frac{1 \pm A_\tau}{2}) \qquad (2.25)$$

for the identification and isolation of dynamical mechanisms that are respectively even or odd under the $A_\tau$ operator. Since it is clear that the only eigenvalues of $A_\tau$ are $\pm 1$, these are idempotent projection operators that obey

$$\begin{aligned} I &= P_A^+ + P_A^- \\ (P_A^\pm)^2 &= P_A^\pm \end{aligned} \qquad (2.26)$$

and thus create a superselection principle for the dynamics associated with parity conserving single-spin asymmetries. When a component of a scattering amplitude is found to be odd under $A_\tau$, the modulus squared of that component will also appear in the expression for the cross section. It is useful to briefly summarize the distinction between the representation of a single-spin asymmetry in the helicity amplitude formalism with the representation formed from $A_\tau$-projected amplitudes. This will be a simplified discussion, considering only one spin-1/2 particle. Let N denote a helicity non-flip amplitude and F denote a helicity flip amplitude;

$$2d\sigma_R = K\{|N|^2 + |F|^2 + 2\,\text{Im}(F^*N)\} \qquad (2.27)$$
$$2d\sigma_L = K\{|N|^2 + |F|^2 - 2\,\text{Im}(F^*N)\} \qquad (2.28)$$
$$\begin{aligned} d\sigma_o &= K(|N|^2 + |F|^2) \\ A_N d\sigma &= -2K\,\text{Im}(F^*N) \end{aligned} \qquad (2.29)$$

In the helicity basis, therefore, the signal for a single-spin asymmetry involves the phase difference between a helicity-flip amplitude and a helicity nonflip amplitude. Using the projection operators for $A_\tau$ given in (2.20)

$$M = (P_A^+ + P_A^-)M = M^+ + M^- \tag{2.30}$$

the same cross sections can be written

$$2d\sigma_R = K(|M^+|^2 - |M^-|^2) \tag{2.31}$$

$$2d\sigma_L = K(|M^+|^2 + |M^-|^2) \tag{2.32}$$

$$d\sigma_o = K|M^+|^2$$
$$A_N d\sigma = K|M^-|^2 \tag{2.33}$$

There are several natural advantages to the use of helicity amplitudes. They have complete rotational invariance, and, using the appropriate spin projections, helicity amplitudes can be extracted from complicated Feynman diagram calculations. However, the requirement in the helicity basis for a single-spin asymmetry—the identification of a phase difference—is notoriously difficult to model effectively except in low-order perturbation theory. The advantages of the transversity basis [12] projected by the operators $P_A^\pm$ are thus specific to the simplifications achieved in the ability to calculate single-spin observables. The most important simplification involves the isolation of the $A_\tau$- odd dynamics. For the hard scattering from a target with polarized protons, the single-spin analyzing power for a jet observable can necessarily be written in terms of the $A_\tau$-odd distributions $\Delta^N G_{q/p\uparrow}(x, k_{TN}; \mu^2), \Delta^N G_{G/p\uparrow}(x, k_{TN}; \mu^2)$ convoluted with $A_\tau$-even spin independent factors. In addition, for the measurement of the polarization of a final-state hadron, all of the spin dependence can be absorbed into the $A_\tau$-odd fragmentation functions $\Delta^N D_{h\uparrow/q}(z, k_{TN}; \mu^2), \Delta^N D_{h\uparrow/G}(z, k_{TN}; \mu^2)$ with $h\uparrow$ representing a polarized proton or lambda hyperon, etc. This isolation of the spin-dependent effects is the basic result first derived, in a more convoluted manner, in ref. 3. The consequences of that isolation include the conclusion that coherent $A_\tau$-odd dynamics in hadronic systems can be studied in hard scattering processes. The discussion here merely uses the simple properties of the $A_\tau$ operator applied to both the hard-scattering component and the soft components of the hard-scattering expansion. Because all of the $A_\tau$-odd dynamics can be absorbed or "factorized" into the $k_T$-dependent distribution or fragmentation functions it is also possible to make a connection between these functions and a sequence of $A_\tau$-odd operators in a twist –expansion [5] [14] in regions where both formalisms are valid.

Although quark spin observables are not as "directly" measurable as those of stable hadrons, much of the preceding discussion can be repeated for asymmetries sensitive transverse quark spin. For light quarks all of the $A_\tau$-odd dynamics can be encapsulated either into the chiral-odd Collins [4] fragmentation function $\Delta^N D_{h/q\uparrow}(z, k_{TN}; \mu^2)$ or into the chiral-odd Boer-Mulders [15] distribution function $\Delta^N G_{q\uparrow/h}(x, k_{TN}; \mu^2)$. These functions appear in the hard-scattering expansion convoluted with $A_\tau$-even factors subject to the additional constraint that one of the other factors must also be chiral odd. Because of their chiral properties, they can be valuable in projecting the transversity distribution. Again, there exists a natural connection with the twist expansion. The four $A_\tau$-odd quantum structures, two classes of fragmentation functions and two classes of distribution functions identified by Mulders and Tangerman [16] involve closely related dynamic origins.

This section has spent some effort discussing the differences between the time reflection operator, T, and the linear operator, $A_\tau$. The author hopes that terms such as "T-odd" applied to transverse spin effects will disappear. We will refer to orbital structures as $A_\tau$-odd. This is much more than just a matter of semantics. We observe that the symmetry here called "artificial time reversal" (a term introduced in Ref. 7) can also be designated "naïve time reversal" [17]. This alternative designation would serve as well. Other terms that more closely identify the behavior of spin-orbit dynamics with the transversity amplitudes of Moravcik and Goldstein [12] such as "transverse parity odd" or "transversity odd" would also be appropriate. Among the many problems that the confusing designation "T-odd" has created is that it obscures the correct application of time reversal invariance in scattering theory to facilitate the discussion of the crossing relations between the $A_\tau$-odd distribution functions and fragmentation functions discussed above. These functions all share a common dynamical origin associated with coherent spin orbit effect in QCD that deserves more theoretical attention. In addition, the incorrect language has obscured the crucial importance of other dynamic symmetries. Within the sector of $A_\tau$-odd dynamics, parity is necessarily even and, hence, by the CPT theorem, CT=+. In the light-quark sector of QCD, isospin also provides important dynamical constraints. The combination of isospin and charge conjugation (G-parity) provides a valuable tool in describing spin-orbit correlations. Another problem created by the incorrect terminology involves the frequent misidentification of the phase difference that occurs in the helicity amplitude description of single-spin asymmetries given in (2.29) with the CP-violating phases in quark mass matrices that lead to true violations of time reversal invariance.

## III. The Construction of $\Delta^N G^{front}$ from Rotating Constituents.

The abstract of Ref. 3 includes the sentence, "It seems convenient to represent the coherent spin-orbit forces in a polarized proton by defining an asymmetry in the transverse-momentum distribution of the fundamental constituents." The parameterization presented there introduces the function

$$\Delta^N G_{q/p\uparrow}(x, k_{TN}; \mu^2) = G_{q/p\uparrow}(x, k_{TN}; \mu^2) - G_{q/p\downarrow}(x, k_{TN}; \mu^2) \quad (3.1)$$

that has been discussed here in terms of its transformation under the operator $A_\tau$. Reference 3 does not explain directly how an orbiting quark can generate this parton density and we will remedy that omission here by giving a simple and direct construction. In Eq. 3.1, $k_{TN}$ is the projection of quark momentum in the direction normal to both the proton spin orientation and the 3-momentum of the proton. This notation was introduced in ref. 3 and will be retained for this paper. With the proton polarized in the $\hat{y}$ direction and moving with momentum in the $\hat{z}$ direction, this identifies the $\hat{x}$ component of quark momentum. Following the Trento conventions [18] this asymmetry can be related to the function introduced by Anselmino, D'Alesio and Murgia [19]

$$\Delta^N G_{q/p\uparrow}(x, k_{TN}; \mu^2) = \Delta^N f_{q/p\uparrow}(x, k_T^2; \mu^2) \quad (3.2)$$

It can also be related to the, now more familiar, nomenclature of Mulders and Tangerman [16]

$$\Delta^N G_{q/p\uparrow}(x, k_{TN}; \mu^2) = -2 \frac{k_{TN}}{M_p} f_{1T}^q(x, k_T^2; \mu^2) \quad (3.3)$$

where $M_p$ is the proton mass. The conventions for $A_\tau$-odd function, both in normalization and in sign, are important. Ref. 18 therefore provides an invaluable guide to checking on the relationships between the theoretical conventions and the experimental asymmetries. However, as will be seen in our discussion of hard scattering from an orbital structure, Eq's. (3.1)-(3.3) are not yet complete. Because an orbital structure is two-valued, the segment of the orbit that contributes must be identified. If we let $\hat{P}$ denote a unit vector along the proton's momentum, it can be seen that equations (3.1)-(3.3) lack the degree of specification needed to identify the value of the vector product, $\hat{\sigma} \cdot (\hat{P} \times \hat{k}_T) = \sin\phi$, that determines the angular segment of the closed orbit at which the scattering occurs. There are two categories of identification that can resolve this indeterminacy:

    1. A specific experimental designation

2. An intrinsic geometrical definition.

From this point, we will include a supplemental label on all expressions for $\Delta^N G$. The construction in this section will produce a distribution $\Delta^N G_{q/p\uparrow}^{front}(x, k_{TN}; \mu^2)$ that describes a quark density in the front of a proton as "viewed" by an oncoming beam. This provides a geometrical definition as explained further in the caption for Fig. 2. A model for a specific scattering process, on the other hand, would provide an experimental specification, such as $\Delta^N G_{q/p\uparrow}^{(DY)}(x, k_{TN}; \mu^2)$ indicating that the distribution is to be "measured" by the Drell-Yan process. The omission of this label in ref. 3 implied an absence of process dependence. This cannot be true. We will try to be very explicit about these conventions as we proceed. The discussion of the connection of our nomenclature with the "gauge-link" specification of these functions introduced in Ref. 10 will be deferred to another paper. The reader need only be aware that a "gauge-link description" can provide either type of designation mentioned above.

Recent theoretical treatment of these quark distribution functions has often focused on what might be termed spectator models.[8] [20] These are models for a specific scattering process, in which a hadron is described in terms of its constituents and amplitudes are generated allowing for gluon exchange or other interactions between the scattered quark and one of the spectator constituents. Even though these are "theoretical" models, they provide the basis for an "experimental" designation of $\Delta^N G$ as defined above. Spectator models have provided an important tool to demonstrate concretely that both orbital angular momentum and initial- or final-state interactions are necessary to generate observable asymmetries. They can also be used to develop an understanding of the origin of spin-orbit correlations. It is appropiate to mention here the model introduced by Burkardt[9]. That paper develops the connection between generalized parton distributions[21] and $k_T$-dependent distributions in the framework of the impact parameter representation. The $A_\tau$-odd distribution (3.1) in his approach is generated by an x-dependent displacement of the quark in impact-parameter space combined with an "attractive" final-state interaction. Burkardt's model is specific to semi-inclusive deep inelastic scattering and, hence, can also be classified as an experimental designation. The "geometrical" approach presented here is complementary to both of these model types. It follows more closely the original suggestion in Ref. 3 by concentrating directly on the kinematics of a hard-scattering event involving an orbiting constituent. In this manner we describe an orbital distribution in terms of an intrinsic property of the proton. We will be able to show later how some of the features of these other models appear directly in our construction. The starting point for this simple construction is indicated in Fig. 2. In the rest frame of a proton polarized in the $+\hat{y}$ direction, we consider the $A_\tau$-odd component of the quark number density projected onto a localized region of the $\hat{x} - \hat{z}$ plane describing a rotating quark with $\hat{L} \cdot \hat{\sigma} = +1$. Using the operator $P_A^-$ defined in eq. (2.25) we get

$$dN_A = P_A^- |\Psi_q|^2 \frac{d^3p}{E} = |\Psi_q|_A^2 \frac{2\pi}{\omega} k_r dk_r dk_y d\phi \tag{3.4}$$

In a region of the $\hat{x} - \hat{z}$ plane containing $r = R_o$ and $\phi$ we simplify further by assuming that $|\Psi_q|_A^2$ is sharply peaked around $k_r = k_o$ with $k_o R_o = 1$ so that after integration over $dk_r$ and $dk_y$ we can write

$$dN_A = \frac{N_A^q}{2\pi} d\phi \tag{3.5}$$

normalized to give

$$\int dN_A = \langle L_y^q \rangle. \tag{3.6}$$

Note that continuity and symmetry provide that $dN_A$ is independent of $\phi$. This construction provides a simple geometrical interpretation of the Bjorken-x variable. To see this we note that for a segment of the orbit as indicated in Fig. 2, we have the 4-momentum

$$k_\mu = (\omega, -k_o \sin\phi, 0, -k_0 \cos\phi)$$
$$\omega = (m_q^2 + \langle k_y^2 \rangle + k_o^2)^{\frac{1}{2}} \tag{3.7}$$

for a rotating quark. We can project this 4-momentum onto light-cone coordinates to give

$$k_+ = \omega - k_o \cos\phi$$
$$k_- = \omega + k_o \cos\phi \tag{3.8}$$
$$k_{TN} = -k_o \sin\phi$$

Following the Trento conventions for a polarized target with the beam particle momentum in the $+\hat{z}$ direction we can therefore specify the target orbit parameters by

$$x = \frac{k_-}{M_p} = x_o + x' \cos\phi$$
$$k_{TN}(x) = -M_p x' \sin\phi \tag{3.9}$$
$$k_{TN}(x) = -M_p [x'^2 - (x - x_o)^2]^{\frac{1}{2}}$$

with

$$x_o = \omega/M_p$$
$$x' = k_o/M_p. \qquad (3.10)$$

With the beam particle directed in the $+\hat{z}$ direction, the front portion of the orbit pictured in Fig. 2 is defined by the arc $\phi \in (0, \pi)$. This construction restricted to the front portion of the orbit therefore gives

$$\Delta_N G^{front}_{q/p\uparrow}(x, k_{TN}(x); \mu^2) = -\frac{N_A^q}{4} x' \sin\phi = -\frac{N_A^q}{4}\left[x'^2 - (x-x_o)^2\right]^{\frac{1}{2}} \qquad (3.11)$$

In this construction, the normalization of $\Delta_N G^{front}_{q/p\uparrow}$ is therefore given by

$$\int_0^1 dx \Delta_N G^{front}_{q/p\uparrow}(x, k_{TN}(x); \mu^2) = -\frac{N_A^q}{4}\int_1^{-1} d(\cos\phi) = \langle L_y^q \rangle/2. \qquad (3.12)$$

Note that this construction does <u>not</u> produce a general $k_T$-dependent quark distribution function. The reason for this is that a stable orbital structure oriented by the proton's spin direction is a highly constrained quantum system. Because of the requirements of orbit continuity, $k_{TN} = k_{TN}(x)$ is not an independent variable. The quark density defined in (3.4) is a local gauge-invariant number density in which momentum fluctuations, $\delta k_y^2$ and $\delta(k_r - k_o)^2$ have already been integrated over. Except for orbital kinematics and the requirement for an orbit-sustaining, confining force directed toward $-\hat{r}$ implicit in the connection with $L_y^q$, the distribution (3.12) has similar locality properties to a collinear quark distribution. In the Mulders-Tangerman[16] formulation, the corresponding $A_\tau$-odd distribution is defined in terms of a nonlocal quark correlator in the light-cone gauge. In addition to an explicit factor of $k_{TN}$ a further functional dependence on $k_T^2 = k_y^2 + k_{TN}^2$ is included in their approach. However, no relationship between $k_{TN}$ and x is specified as this is assumed to reside within the quark correlator. The advantages of the more formal nonlocal approach occur in the systematic connection to other $k_T$-dependent distributions. A possible advantage of the alternate construction presented here resides in the simplicity of the normalization (3.12) and the concrete geometric picture for a local gauge-invariant quark density that the $A_\tau$-odd projection makes possible. The two different formulations of the underlying distribution allow for significant cross-checks.

From Eq. (3.8) we see that the "back" portion of the proton defined by the arc segment $\phi \in (\pi, 2\pi)$ differs in the sign of $k_{TN}(x)$ so that we have

$$\Delta_N G_{q/p\uparrow}^{back}(x, k_{TN}(x); \mu^2) = -\Delta_N G_{q/p\uparrow}^{front}(x, k_{TN}(x); \mu^2) \tag{3.13}$$

This observation will be used extensively in the next section where we describe how "soft" initial-and/or final-state interactions combined with the "hard" scattering from a rotating constituent can lead to an $A_\tau$-odd observable asymmetry. The point that our construction makes clear is that it is necessary to "unfold" the effect of these soft interactions using a space-time picture of the scattering process and, at some point, get to an observable related to $\Delta_N G_{q/p\uparrow}^{front}$, in order to relate experimental symmetries to a measurement of $\langle L_y^q \rangle$. Fig. 3 presents a representation of the properties of an orbital quark in a Lorentz-boosted frame that can serve as a memory aid for the features discussed in this section.

**IV. Oriented and Non-Oriented Soft Interactions in the Initial or Final State.**

The basic framework of the QCD "hard-scattering" model is based on the observation that a large momentum transfer can necessarily be associated with scattering from a pointlike constituent. Because the hard-scattering event is localized spatially, it is possible to initially neglect some interference effects and, for a large class of observables, this leads to a factorization property[22] wherein experimental cross sections can be written as a convolution over internal kinematic variables of a perturbatively-calculated cross section with hadronic distribution and/or fragmentation functions. This fundamental approach has been justified by an extensive theoretical framework [23] and has been validated by systematic experimental tests. [24] An important factorization property that has been established by this program involves the "universality" of a set of hadronic distribution and fragmentation functions that embody the nonperturbative component of the "model". This means that the $k_T$-integrated, spin-averaged quark distribution $G_{q/p}(x; \mu^2)$ defined within a specified factorization prescription can be "measured" in one process and then used to calculate other observables. Within a given prescription, the dependence of the distribution on the factorization scale itself can be calculated perturbatively. Therefore, the formulation of the model allows for calculations that can be systematically improved order by order in perturbation theory.

The description of single-spin observables associated with the localized hard scattering from an orbiting constituent fits into a "borderline" category of this model. The basic assumption, that the large-momentum transfer occurs in the localized scattering from a pointlike constituent, remains intact. The reason that this assumption does not lead to a "universal" distribution that is measured to be the same in all experimental processes follows from the properties of an orbital distribution as discussed in Sec. III. Because of orbital symmetry, observable spin asymmetries can exist only if there are soft initial- and/or final-state interactions in addition to the localized hard scattering. These additional interactions are necessary to prevent the cancellations that would otherwise occur between segments of the closed orbital motion with opposing $k_{TN}$. On consideration, it is clear that these interactions must also appear in spin-averaged cross sections where they either cancel or involve fluctuations that are absorbed into the definitions of $A_\tau$-even distributions. The interpretation of the hard-scattering model as the representation of a local event requires that we account for these additional interactions by considering the average of two potential hard-scattering events with different kinematics. In general, this description is not "universal" as defined above but it is definitely factorizable. It is important to keep in mind that a <u>new</u> factorization property applies to a description of these measurements. This very strong factorization property occurs because all hard-scattering processes involving light quarks are necessarily even under the $A_\tau$ operator as shown in Sec. II. All the $A_\tau$-odd dynamics can therefore be included in an "effective" $k_{TN}$-dependent orbital distribution. This effective distribution can be related to the intrinsic distribution $\Delta^N G_{q/p\uparrow}^{front}(x, k_{TN}; \mu^2)$ constructed in Sec. III. However, since different combinations of the soft interactions can preferentially expose different segments of the constituent orbit, the approach can lead to significant process dependence for the experimentally-designated distributions defined there. We will demonstrate this process dependence by directly examining the effect of soft interactions in three different scattering processes. In the cases we consider, the nature of the interactions allow for straightforward phenomenological estimates of the impact on the kinematic convolutions of the hard-scattering model. This enables a quantitative connection between the experimentally-designated distributions and the underlying "intrinsic" orbital structure. An independent discussion of the formal factorization properties found in $k_T$-dependent distributions based on the nonlocal correlator can be found in Ref.[25].

The factorization properties inferred from the arguments above are illustrated in Fig. 4 for three different experimental asymmetries: $A_N d\sigma(ep\uparrow \to eqX)$ (SIDIS), $A_N d\sigma(\bar{p}p\uparrow \to e\bar{e}X)$ (DY) and $A_N d\sigma(pp\uparrow \to qX)$ (Gq). For the application of the hard-scattering model, the indicated "soft" exchanges in these diagrams can be separated into two categories:

1. non-oriented interactions,
2. spin-oriented interactions.

We will deal with the effects of the two categories separately since there impact on the kinematic convolutions in the model can be separated.. The non-oriented interactions in the scattering process can produce an effective screening by the mechanism of jet energy loss. Jet energy loss involves the transfer of momentum from a high-energy constituent with SU(3) color charge to a hadronic medium. There exists considerable evidence from the di-jet events in proton-proton and proton-nucleus collisions for jet energy loss.[26] This energy loss has been characterized by Brodsky and Hoyer[27] who considered the interior of a hadron to be a color-polarizable medium. In analogy to the passage of an electron through a dielectric medium, soft multiple collisions of a fast-moving quark or gluon generate gluon bremmstrahlung to deplete the energy of a jet. Brodsky and Hoyer give the expression for jet energy loss in the form

$$\langle dE/dl \rangle \leq \frac{1}{2} \langle k_T^2 \rangle \tag{4.1}$$

where l is the path length and $\langle k_T^2 \rangle$ is the mean-squared transverse momentum generated by the class of soft scatterings. Just as in the Landau-Pomeranchuk –Migdal effect [28] the formation zone of the emitted gluon radiation cannot be larger than the nuclear or hadronic size.

A slightly different approach to jet energy loss can be found in Ref. [29] where soft color exchanges involving a fast quark or gluon creates a web of color flux that causes the energetic parton to decelerate. The expression for the energy loss in this approach is

$$\langle dE/dl \rangle \cong \sigma \tag{4.2}$$

where $\sigma$ is the effective chromoelectric string tension for color flux tubes. The string tension has the approximate quantitative value

$$\sigma \cong 0.2 Gev^2$$
$$\sigma \cong 1.0 Gev/fm \tag{4.3}$$

These two phenomenological approaches to SU(3) color acceleration can be shown to give a similar space-time picture and to have similar formation-zone constraints. Data from a variety of nuclear targets [30][31] seem to require a higher value of jet energy loss than is given in (4.2) and (4.3). These data can be explained in terms of higher representations of color charges such as gluons and/or multiple flux tubes. The important aspect of this discussion for the study of single-spin asymmetries from orbital structures is that an incoming (or outgoing) constituent with SU(3) color charge can experience a mean energy loss in excess of 1 Gev/fm while traveling from one side of the proton to another. With a proton charge radius of

$$\langle R_p^2 \rangle^{\frac{1}{2}} \cong 0.8 fm$$

we see that jet energy loss can provide a significant kinematic selection in the hard-scattering model for scattering events involving orbital constituents. From the phenomenological point of view it is important to note that the effect of jet energy loss is always dissipative. For initial-state interactions, it favors a scattering event in the "front" part of the orbit. For final-state interactions, it favors a scattering event detector or "back" side of the orbit. The impact of this kinematic selection can be represented in a simple geometric fashion.

Using an eikonal (straight-line) approximation for the energy loss of a color-charged constituent we can parameterize the effect of initial-state interactions in the form

$$\Delta^N G_{q/p\uparrow}^{ISI}(x, k_{TN}; \mu^2) = \Delta^N G_{q/p\uparrow}^{front}(x, k_{TN}; \mu^2) + \eta_I(x) \Delta^N G_{q/p\uparrow}^{back}(x, k_{TN}; \mu^2)$$
$$= (1 - \eta_I(x)) \Delta^N G_{q/p\uparrow}^{front}(x, k_{TN}; \mu^2) \qquad (4.4)$$

where $\eta_I(x)$ parameterizes the indicated screening effect. Similarly, the impact of non-oriented soft interactions in the final state can be parameterized by

$$\Delta^N G_{q/p\uparrow}^{FSI}(x, k_{TN}; \mu^2) = \eta_F(x) \Delta^N G_{q/p\uparrow}^{front}(x, k_{TN}; \mu^2) + \Delta^N G_{q/p\uparrow}^{back}(x, k_{TN}; \mu^2)$$
$$= (1 - \eta_F(x)) \Delta^N G_{q/p\uparrow}^{back}(x, k_{TN}; \mu^2) \qquad (4.5)$$
$$= -(1 - \eta_F(x)) \Delta^N G_{q/p\uparrow}^{front}(x, k_{TN}; \mu^2)$$

The general form of the impact on non-oriented interactions including both initial-state and final-state interactions can then be given as

$$\Delta^N G_{q/p\uparrow}^{non-oriented}(x, k_{TN}; \mu^2) = (\eta_F(x) - \eta_I(x)) \Delta^N G_{q/p\uparrow}^{front}(x, k_{TN}; \mu^2) \qquad (4.6)$$

Phenomenological estimates of the screening parameters can be obtained using (4.1)-(4.4). However, we will not do so here. At this point we will simply mention some of the extreme limits of the phenomenological screening parameters used above. The simple parton model neglects all screening effects and we can write

$$\eta_I^{parton}(x) = \eta_F^{parton}(x) = 1 \qquad (4.7)$$

and in this approximation the expressions (4.4), (4.5) and (4.6) all give zero. This demonstrates the idea that the interactions are necessary to produce nonzero asymmetries. Another useful approximation is given by the "black-body" limit

$$\eta_I^{black}(x) = \eta_F^{black}(x) = 0 \qquad (4.8)$$

In the black body limit, expressions (4.4) and (4.5) both give a direct measurement proportional to orbital angular momentum while (4.6) vanishes. Both these limits have direct classical interpretations that can be demonstrated by simple table-top experiments. These demonstrations reinforce the fact that the basic mechanism behind single-spin asymmetries in hard-scattering processes is not at all mysterious. The understanding of the kinematics allows the focus to remain on the question of what these observables can uncover about spin-orbit effects in proton structure. In addition, it is important to keep in mind the geometrical constraints

$$\lim_{x \to x_{max}} \eta_{I,F}(x) = 1 \tag{4.9}$$

so that the large-x behavior of (4.4) to (4.7) is always such that the asymmetries vanish more rapidly than $k_{TN}(x)$. We will return to say more on the subject of screening after considering the effect of spin-oriented soft interactions.

Spin-oriented components for the soft-momentum transfers involved in the initial-state or final-state interactions are possible because the orbiting constituent is not a free particle but, instead, is being accelerated by an attractive force

$$\frac{dk_i}{dt} = -\kappa r_i \tag{4.10}$$

It is necessary to account for this force in any description of the scattering of the orbiting particle. According to the spatial geometry of the orbital structure defined in Sec. III, a quark with $x \geq x_o$ orbiting with $\hat{L} \cdot \hat{\sigma} = +1$ in a proton polarized in the $+\hat{y}$ direction is on the "left" side of the proton as viewed from the beam. In this configuration, the confining force therefore has a component in the $-k_{TN}$ direction. After scattering via hard-photon exchange, the quark remains subject to this confining force. The space-time progression of this scattering process is sketched in Fig. 5 which demonstrates that a spin-directed final state interaction exists as the quark emerges from the proton. This final-state interaction is capable of generating an average spin-oriented momentum transfer, $\delta k_{TN}$, before the hadronization process is complete. In this case the average momentum transfer generated by this mechanism is in the $-k_{TN}$ direction. This feature of an orbital structure has been labeled "chromodynamic lensing" by Burkardt [9] and plays a dominant role in generating the spin asymmetry for semi-inclusive deep inelastic scattering in his model.

Spin-oriented momentum transfers can also occur in annihilation processes. The sketches in Fig. 6 show the comparable space-time progression for the annihilation of an orbital quark at large x by an incoming antiquark to produce a virtual photon in the Drell-Yan process. Here, the annihilation of color and charge can "release" a directed momentum in the $+k_{TN}$ direction, opposite to the direction of acceleration of the orbital quark before annihilation.

The comparison between the initial-state interactions in the Drell-Yan process and the final-state interactions in semi-inclusive deep inelastic scattering leads a geometric and dynamic understanding of the Collins conjugation relation. In Ref. [10] Collins found the relationship

$$\Delta^N G^{DY}_{q/p\uparrow}(x,k_{TN};\mu^2) = -\Delta^N G^{SIDIS}_{q/p\uparrow}(x,k_{TN};\mu^2) \qquad (4.11)$$

using the gauge link formalism to relate the nonlocal quark correlators probed in the two different processes. In our construction of a local orbital density, the same result occurs from a combination of the screening arguments given in (4.4) and (4.5) using SU(3) color charge conjugation and an application of the same color charge conjugation arguments to the spin-directed momentum-transfers to get

$$\delta k^{DY}_{TN} = -\delta k^{SIDIS}_{TN} \qquad (4.12)$$

for the two processes. The fact that our localized "hard-scattering" plus soft interactions approach can replicate Collins' result is quite significant. The conjugation relationship (4.11) does not depend on the relative amount of screening and directed momentum transfer, only on the requirement that both mechanisms involve QCD. It illustrates that the non-local $A_\tau$-odd quark correlator describes <u>all</u> of the nonperturbative dynamics in these processes—including spin-directed binding effects and the impact of the initial- or final-state interactions. The comparison validates the gauge formulation of QCD even in regimes where nonperturbative effects are dominant. F. Pijlman [32] has noted that the path integral approach to the nonlocal quark correlator places the calculation of single-spin asymmetries in a formal analogy to the calculation of the Aharonov Bohm [33] asymmetry. The Aharonov Bohm asymmetry tests QED gauge invariance in spatial regions where the electromagnetic field-strength tensor vanishes while Collins conjugation tests gauge invariance for QCD in kinematic regions where the effective degrees of freedom can quite different from the perturbative formulation. An experimental test of the Collins relationship can therefore provide a strong probe of the formulation of QCD as a local gauge theory. The observation of a violation of this conjugation relation would provide an indication that QCD, by itself, is not adequate to describe the complex phenomenology of hadronic physics.

The color structure of the hard-scattering cross section also plays and important role in the understanding of the process dependence associated with the presence of soft interactions. The oriented initial- and final-state interactions are not always present! In the spatial description of the spin asymmetry $A_N d\sigma(pp\uparrow \to jetX)$, the exchange of a hard gluon in the t-channel liberates the orbital quark from the spin-directed confining force in (4.10). This leads to the prediction that in this purely hadronic process, in the appropriate kinematic region, the asymmetry can be affected only by non-oriented initial- and final-state interactions leading to screening effects. This is significantly different than the "hard photon" processes discussed above. The role of color-flow in these $A_\tau$-odd spin asymmetries reflects the analogous "flux-drag" effect observed in the angular structure of multiplicities in $e^+e^- \to \bar{q}qG$ compared to $e^+e^- \to \bar{q}q\gamma$ events.[34] [35]. Instead of seeing the effect of different QCD flux configurations in particle multiplicities, we can observe them in the process dependence of measurements of single-spin asymmetries. The importance of the color structure of the hard-scattering mechanism in the calculation of asymmetries helps define the distinction between factorization and universality discussed above. In perturbative QCD, it is typical that more than one color flow pattern is involved in the amplitude for a hard subprocess. The $A_\tau$-odd nature of single-spin asymmetries and the distinction between oriented and non-oriented soft interactions may allow for an experimental separation of color-weighted cross sections in the manner of that originally postulated by Ellis, Marchesini and Webber [36]. Having different experimental probes can provide a strategy for the extraction of the intrinsic distribution $\Delta^N G_{q/p\uparrow}^{front}(x, k_{TN}; \mu^2)$, defined in Sec. III. The hadronic process $A_N d\sigma(pp\uparrow \to jetX)$ can provide an opportunity to measure inclusive screening parameters in certain kinematic regions with great accuracy and without the complication of spin-directed effects.

This short discussion illustrates the process dependence inherent in the calculation of spin asymmetries. The process dependence complicates the goal of relating measurements in different processes to a distribution that can be normalized to $\langle L_y^q \rangle$ or to $\langle L_y^G \rangle$. However, quantitative phenomenological estimates can be made for these soft deflections. The possibility of studying spin asymmetries in 2-jet events [37] or in multijet events opens the door on more complicated color-flow configurations. It is not clear that the tools described here are adequate for all such situations.

### V. Hard Scattering and Single-Spin Asymmetries

This paper has taken a direct, primitive approach to the description of $A_\tau$-odd spin asymmetries occurring in hard-scattering processes. By describing a local scattering from an orbital structure we attempt to relate the kinematic effects leading to observable experimental asymmetries to an intrinsic distribution, $\Delta^N G_{q/p\uparrow}^{front}(x, k_{TN}; \mu^2)$, that, by construction, is normalized in (3.12) to give a measure $\langle L_y^q \rangle$. As discussed by Biedenharn and Louck [38], any local measurements of orbital angular momenta are constrained by indeterminacy relations dictated by the commutation relations

$$\begin{aligned}\left[L_y, \cos\phi\right] &= i\sin\phi \\ \left[L_y, \sin\phi\right] &= -i\cos\phi\end{aligned} \quad (5.1)$$

In inclusive scattering experiments on a polarized proton, these commutators can be reconciled with local observables by treating $L_y$ as a quantity with virtual fluctuations and allowing for a distribution of possible values. The mean value $\langle L_y^q \rangle$ represents a property of this distribution that is accessible to measurement. In this sense, our approach follows the spirit of the suggestion by Lurcat [39] concerning orbital observables in composite systems. The construction described in Sec. III and the normalization (3.12) is therefore valid provided that, for each specific flavor of quark

$$\left|\langle L_y^q \rangle\right| << \frac{1}{2} \tag{5.2}$$

Spectator models [8],[20] and other approaches to describing orbital structures [9] have their own constraints that represent these indeterminacy relations. For example, the process of light-cone quantization involving Fock states of definite $L_z$ [40],[41] is compatible with these constraints for measurements involving transverse asymmetries. The formulation is naturally done with helicity amplitudes but (2.27)-(2.33) can be used to connect the two formalisms.

A local description of $A_\tau$-odd dynamics is made possible by the idempotent projection operators $P_A^\pm$ defined in (2.25) and (2.26). This allows for the definition of the intrinsic distribution $\Delta^N G_{q/p\uparrow}^{front}(x, k_{TN}; \mu^2)$ but the program to relate this distribution to experimental spin asymmetries is complicated by the crucial role played in these measurements by nonperturbative initial-state and final-state interactions. This approach is quite different from that involving a nonlocal correlator. Our attempt to focus, in this paper, on the explanation of some fundamental concepts without introducing unnecessary complications rules out a serious attempt at phenomenology but we found that it is possible to explain the significant process-dependence generated by these soft interactions in this approach. Significant phenomenological fits have been made by Anselmino et. al [42] for spin asymmetries in such processes as $pp\uparrow \to \pi X$ and $ep\uparrow \to e\pi X$ and these fits provide a quantitative basis for further progress. In particular, the fact that all $A_\tau$-odd dynamics can be isolated makes it possible to compare the results of our approach with the process-dependence inherent in various twist-3 mechanisms found in collinear factorization. [43] [44] In kinematic regions where both approaches can be valid, this provides a possible strategy to "unmask" the intrinsic distribution.

In the collinear approximation to the hard-scattering model, the parton fluxes that occur in the formulas for cross sections are represented as parton densities times kinematic factors. For $A_\tau$-odd structures this approximation breaks down because of orbital symmetry and initial-state or final-state interactions are required. These additional interactions can be neglected for the calculation of jet observables in most $k_T$-averaged processes. However, Brodsky [45] has shown that such soft interactions must also be included in an understanding of diffractive events in deep inelastic scattering as well as in the description of nuclear shadowing and anti-shadowing. Moreover, the description of rare events in high-energy extrapolations of the hard-scattering model have been shown to be sensitive to the nature of the "underlying – event" in Monte Carlo simulations of processes at measured energies. [46]   Initial and final-state interactions help shape the "underlying-event" and it is hoped that the soft interactions required for observable single-spin asymmetries and also required for the other processes identified by Brodsky can provide an aid in formulating these extrapolations.

Other critical topics intentionally omitted here include a discussion of the dynamical mechanisms leading to orbital structures and a discussion of the many relationships that connect $\Delta^N G_{q/p\uparrow}^{front}$ to the other $A_\tau$-odd functions identified by Mulders and Tangerman.  These are topics that deserve their own forum.

Acknowledgements.  The author has been fortunate to have significant conversations on the topic of this paper with Mauro Anselmino, Stan Brodsky, Matthias Burkardt, John Collins, Leonard  Gamberg , Gary Goldstein, Steve Heppelman, Alan Krisch, Jian-wei Qiu, and Fetze Pijlman.  These gentlemen have strongly influenced the manuscript but it should be clearly proclaimed that the errors and omissions contained herein are solely the author's fault.

Table 1.

|  | $\Sigma_x$ | $\Sigma_y$ | $\Sigma_z$ |
|---|---|---|---|
| Charge conjugation  C | + | + | - |
| Parity  P | - | + | - |
| Time reflection T | - | + | + |
| (CPT) | + | + | + |
| Artificial time reversal $A_\tau$ | + | - | + |
| (P $A_\tau$) | - | - | - |

The combination of rotational invariance plus finite symmetries strongly constrains single spin observables. With the conventions of Fig. 1, the transformation of spin components $(\Sigma_x, \Sigma_y, \Sigma_z)$ under the finite transformations C, P, T and $A_\tau$ are shown in the table. All spin observables are even under CPT and odd under P $A_\tau$.

**Figure Captions**

Fig. 1   The scattering process $12 \to 34$ is shown in the CM system in the top drawing. In the conventions of Jacob and Wick (Reference 6) the scattering takes place in the x-z plane with the normal to the scattering plane defined by $\hat{y} \propto (\hat{k}_1 \times \hat{k}_3)$.

The middle drawing shows the time-reversed process $\overline{34} \to \overline{12}$. The normal to the scattering plane is $\hat{y}_T \propto ((-\hat{k}_3) \times (-\hat{k}_1)) = -\hat{y}$. Time reversal also changes the sign of $\sigma_y$, $((\sigma_y)_T = -\sigma_y)$ and, hence does not change the sign of $\hat{\sigma} \cdot (\hat{k}_1 \times \hat{k}_3)$. The operator $\hat{\sigma} \cdot (\hat{k}_1 \times \hat{k}_3)$ is even under time reversal.

The impact of the $A_\tau$ operator on the scattering process $12 \to 34$ is shown in the bottom figure. The normal to the scattering plane in this situation is $\hat{y}_A \propto ((-\hat{k}_1) \times (-\hat{k}_3)) = \hat{y} = -\hat{y}_T$. The spin operator changes sign under $A_\tau$ and, hence, the operator $\hat{\sigma} \cdot (\hat{k}_1 \times \hat{k}_3)$ is odd under $A_\tau$. The effect of other finite symmetries on spin observables is summarized in Table 1.

Fig. 2   A planar projection of an $L_y = +1$ constituent rotating in a proton. The incoming beam is shown in the $+\hat{z}$ direction. The kinematics of the rotating quark are defined by equations (3.7) and (3.8) in the text. Expressing the momentum in terms of Bjorken x and $k_{TN}(x)$ is done in equations (3.9) and (3.10). The orbit position $\phi$ determines both Bjorken x and $k_{TN}$. The "front" of the proton is defined by the direction of the incoming beam and, for positive $L_y$, has $k_{TN}(x) < 0$. Larger values of Bjorken x ($x > x_o$) correspond to $\cos\phi > 0$ and are on the "left" of the proton as seen from the beam. The local distribution, $\Delta^N G_{q/p\uparrow}^{front}$ contains a factor $k_{TN}(x)$ that ensures that it is odd under $A_\tau$.

Fig. 3   For a proton polarized in the $+\hat{y}$ direction the sketch labeled $L_y = +$ shows an $A_\tau$-odd distribution with $\hat{\sigma} \cdot \hat{L} = +$ with the conventions as defined in Fig. 2. The shaded areas indicate the high-x kinematic region where the momentum of the orbital constituent is "blue-shifted" toward the beam. The sketch labeled $L_y = -$ indicates a distribution with $\hat{\sigma} \cdot \hat{L} = -$. The presence of nonrotating matter is indicated by the dashed boundary. The contribution of the distribution to a spin asymmetry is determined by oriented and non-oriented soft interactions as described in Sec. IV.

Fig. 4   The factorization properties of single-spin asymmetries for the processes $ep\uparrow \to eqX$, $\pi p\uparrow \to e^+e^-X$, and $hp\uparrow \to qX$ as described in the text. Initial-state and final-state interactions determine the net impact of the $A_\tau$-odd distribution.

Fig. 5   A time sequence showing the contribution of the confining force (indicated by a coiled line) on the spin-oriented momentum transfer in the process $ep\uparrow \to eqX$. Scattering at large Bjorken x preferentially occurs on the left of the polarized proton for $\hat{\sigma}\cdot\hat{L}=+$. The confining force leads to $\delta k_{TN}<0$ as shown.

Fig. 6   A time sequence showing the contribution of the confining force (indicated by a coiled line as in Fig. 5) on the spin-oriented momentum transfer for the process $\pi p\uparrow \to e^+e^-X$. At large Bjorken x, the annihilation process leads to a release of momentum, $\delta k_{TN}>0$.

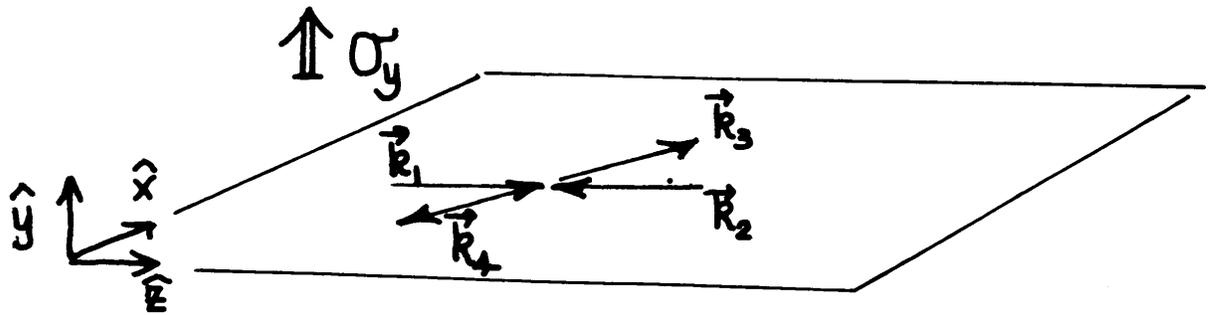

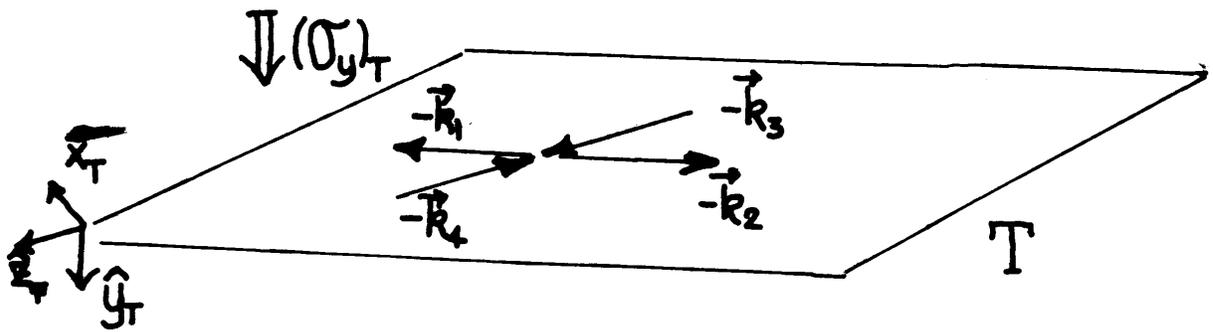

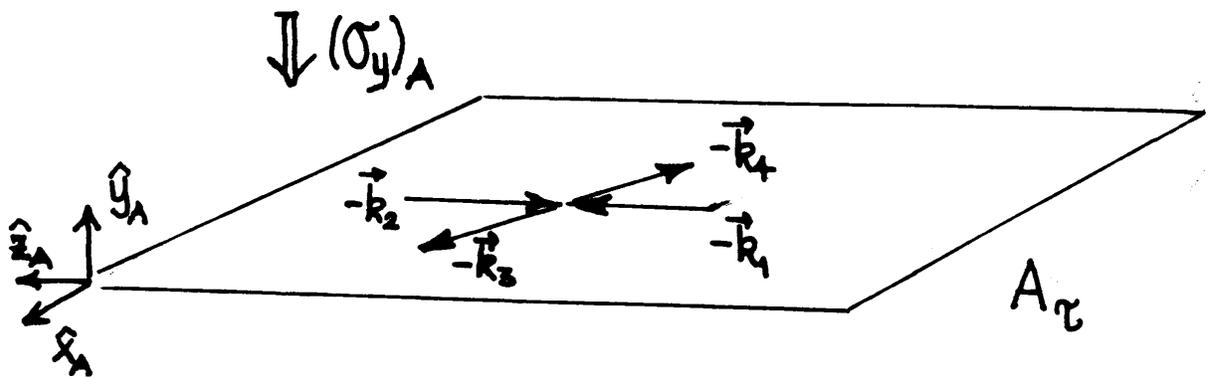

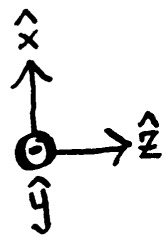

$$L_y = \hat{L} \cdot \hat{S}_p = +$$

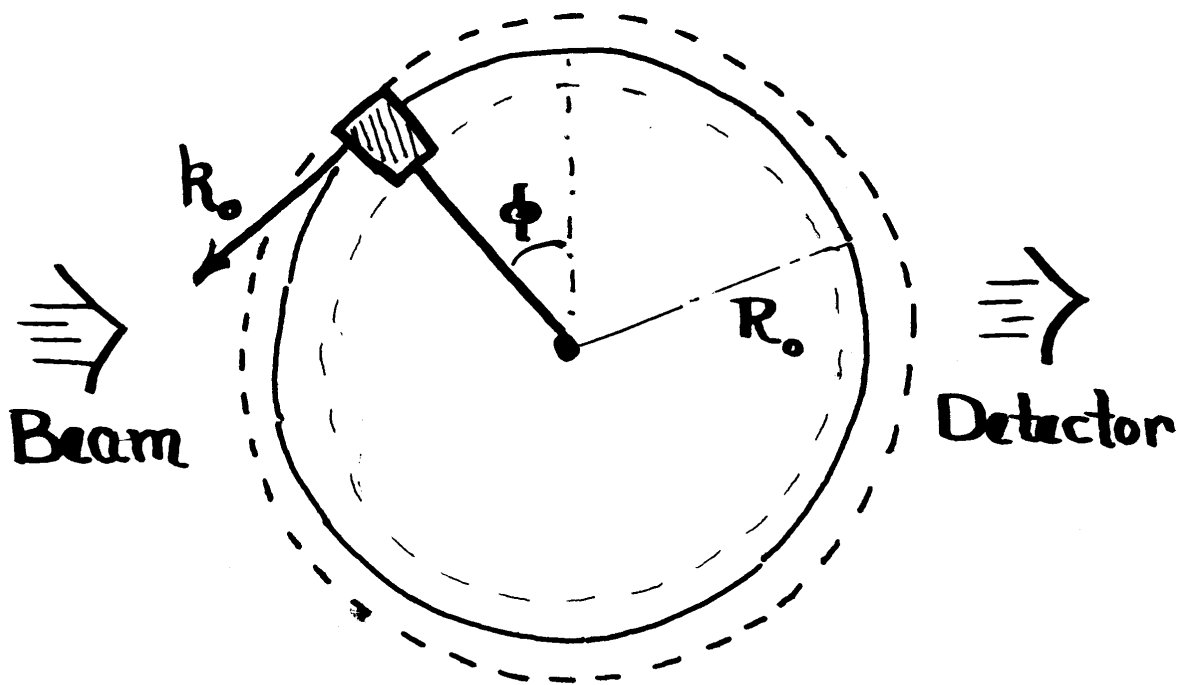

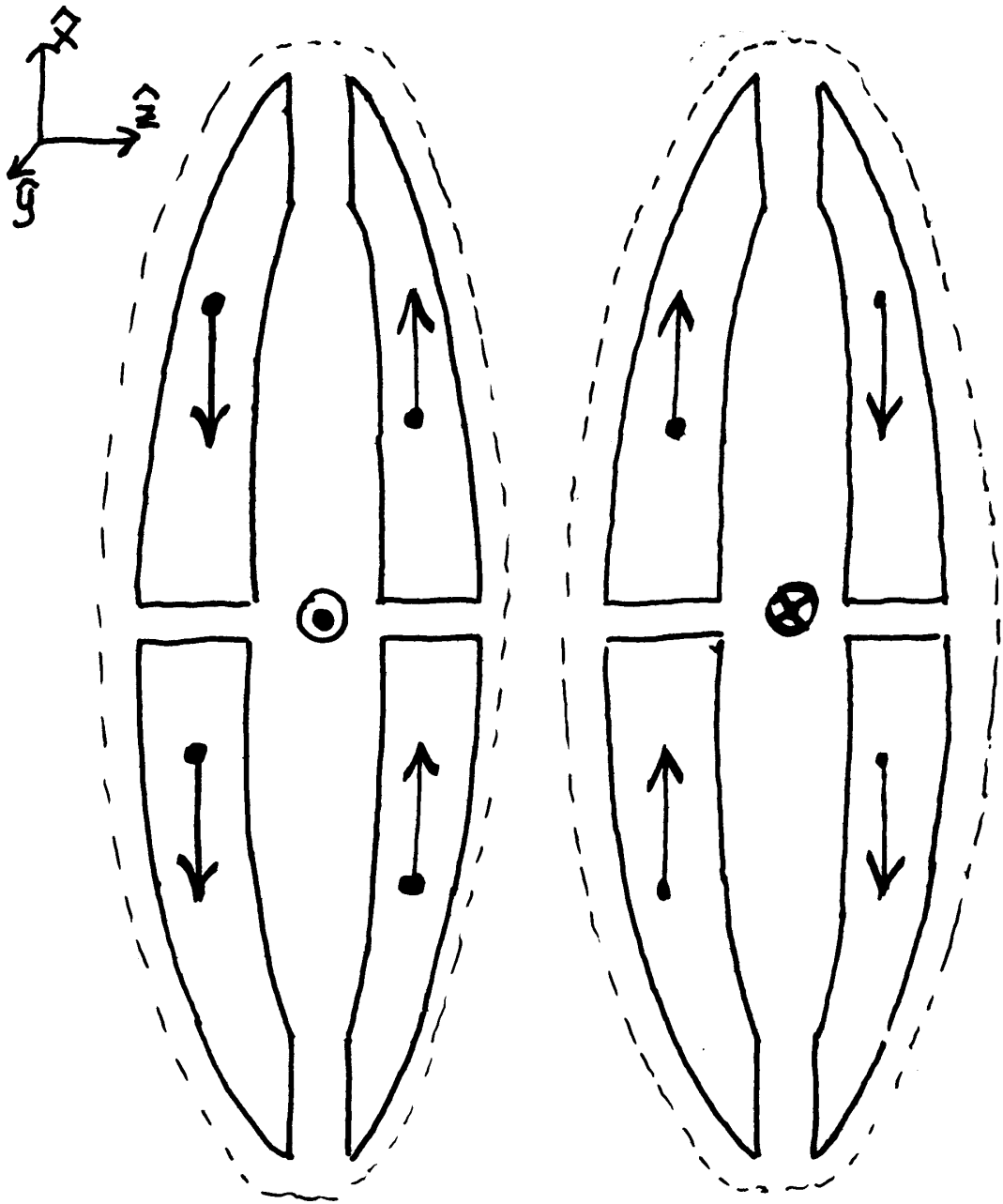

$L_y = +$      $L_y = -$

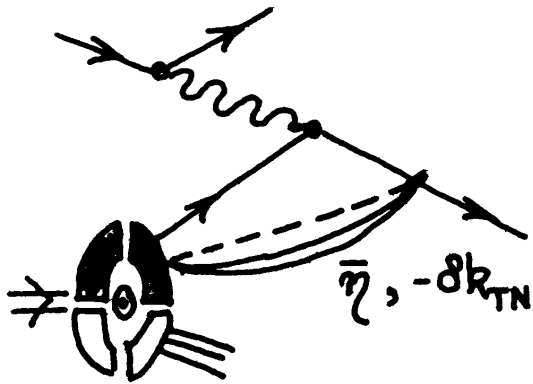

$$ep\uparrow \to eqX$$

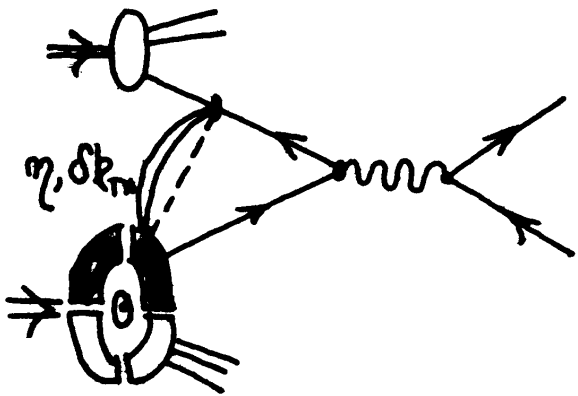

$$\uparrow p\uparrow \to e^+e^-X$$

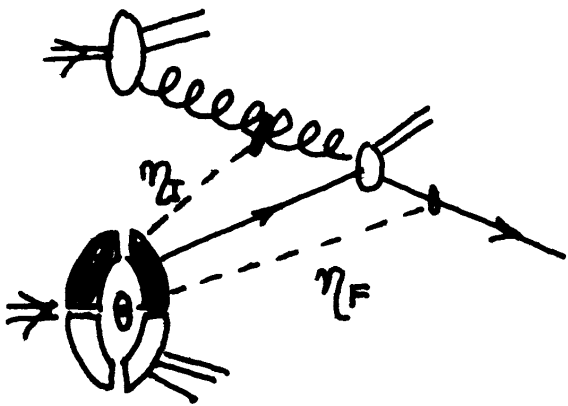

$$hp\uparrow \to qX$$

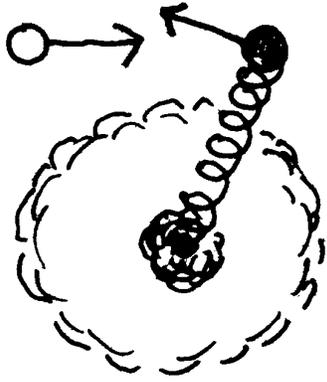

(a)

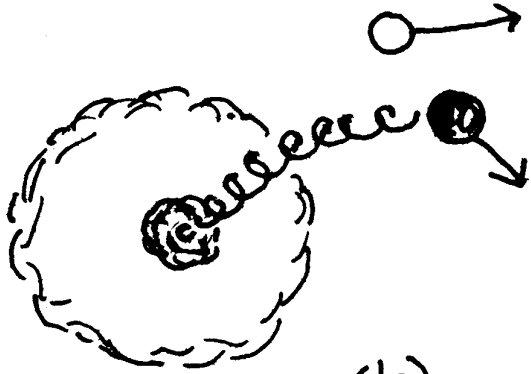

(b)

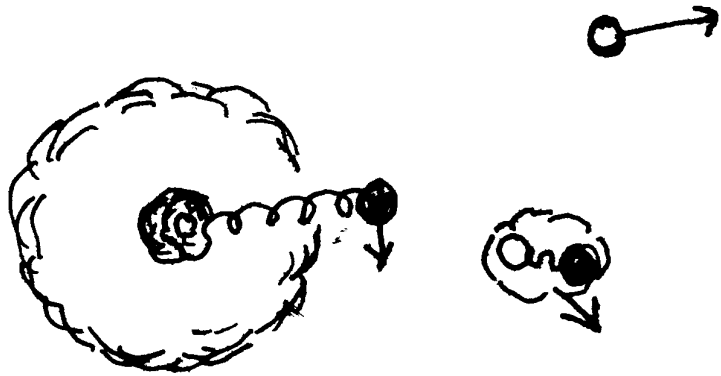

(c)

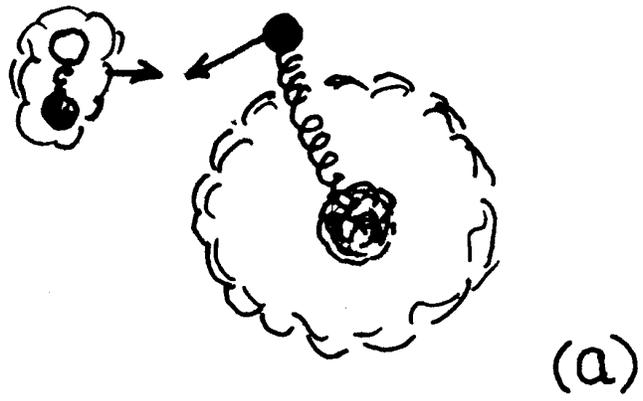

(a)

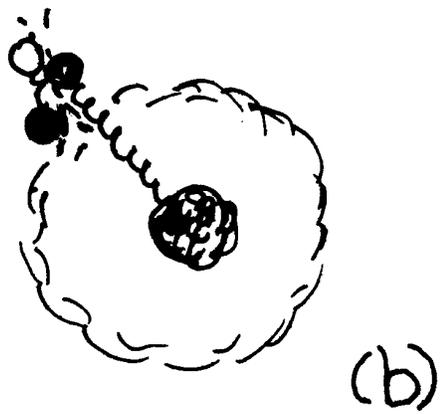

(b)

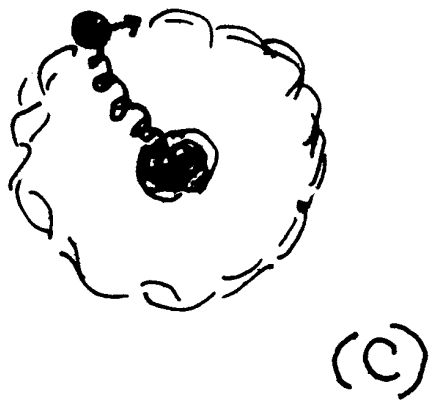

(c)